# Optomechanical coupling in photonic crystal supported nanomechanical waveguides

W.H.P. Pernice[1], Mo Li[1] and Hong X. Tang[1,*]

[1] *Departments of Electrical Engineering, Yale University, New Haven, CT 06511, USA*
*Corresponding author: hong.tang@yale.edu*

**Abstract:** We report enhanced optomechanical coupling by embedding a nano-mechanical beam resonator within an optical race-track resonator. Precise control of the mechanical resonator is achieved by clamping the beam between two low-loss photonic crystal waveguide couplers. The low insertion loss and the rigid mechanical support provided by the couplers yield both high mechanical and optical Q-factors for improved signal quality.

**OCIS codes:** (120.4880) Optomechanics; 130.3120 Integrated optics devices; 350.4238 Nanophotonics and photonic crystals.

## 1. Introduction

Over the past years Silicon nanophotonics has been established as a promising platform for densely integrated optics [1-3]. A multitude of optical components traditionally employed in fiber-based systems have been implemented using silicon nano-waveguides and resonators [4-7]. Because of the high refractive index contrast between silicon and a dielectric substrate or air very compact devices can be realized and potentially be integrated with conventional electronic circuits. Recently it was realized, that such nanophotonic components are size-matched to nanoelectromechanical systems (NEMS) and thus show a route for integrated optomechanical circuits [8]. Instead of relying on electrical methods for device actuation and readout it was demonstrated that it is possible to generate sufficient optical force to set nano-mechanical resonators into motion [9]. Thus all-photonic actuation and signal detection can be realized on a silicon chip.

Previously, optical methods for the detection of nanomechanical motion have relied on interferometric schemes, using Fabry-Perot or Mach-Zehnder configurations. In order to improve the detection sensitivity, devices employing high finesse optical cavities have also been exploited [10-12]. If the mechanical resonator is embedded in a high quality optical cavity, the mechanical displacement of the device modulates the resonance frequency of the cavity. The dispersive coupling of the mechanical resonator and optical cavity enables readout of the mechanical motion by monitoring the optical resonance frequency of the cavity. Compared with interferometric methods, the detection sensitivity is expected to be enhanced by a factor related to the finesse $F$ of the optical cavity.

In this article we propose and experimentally demonstrate an NEMS embedded photonic circuit that provides high Q performance in both optical domain and mechanical domain. The optomechanical signal is therefore significantly enhanced. We realize optical Q enhancement by embedding a nanomechanical beam resonator into an optical race-track resonator. In order to achieve good mechanical support and thus high mechanical quality factors, we design special photonic crystal waveguide couplers that yield rigid mechanical clamping while giving minimal distortion to the optical modes. Thus low transmission loss is maintained and high optical quality factors can be achieved. We demonstrate a mechanical resonator with a quality factor of 1200 in an optical cavity with a Q factor of 4200.

## 2. Design of the optical resonator

We employ a traveling wave resonator – the simplest type is a race track – and use the straight portion of the race-track to form a nanomechanical resonator. In order to achieve high optical Q it is paramount to minimize the propagation loss inside the race-track. The optical quality factor is directly related to the optical loss during one round trip of the optical mode. Silicon nanowire waveguides fabricated by our processes routinely yield propagation losses of 3~5dB/cm. For a ring with radius of 20 µm, the corresponding intrinsic Q-factor is then estimated to be 234000. However, the measured loss will be much higher due to the perturbation of the optical mode by the embedded mechanical device

To embed NEMS device in the optical resonator, a commonly practiced process involves a wet chemical etch to release part of the ring from the substrate. This starts with patterning an etch window using optical lithography and then etching the wafer in buffered oxide etchant. The chemical etching process, however, is highly isotropic and thus will undercut the waveguide beyond the window area defined by the photoresist. Consequently, the resultant length of the released resonator is longer than the defined width of the mask window and difficult to control. This further limits the resonant frequency of the beam one can achieve.

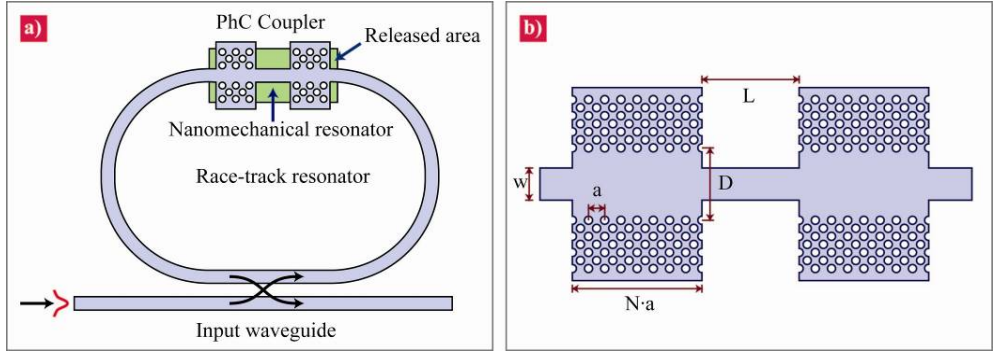

Figure 1: a) The outline of the design to achieve opto-mechanical coupling. We employ a race-track optical resonator to achieve high optical Q. A nanomechanical beam is realized by removing the underlying oxide under parts of the race-track. For good mechanical clamping and precise definition of the beam length the resonator is supported by two photonic crystal couplers. b) The layout of the photonic crystal waveguide coupler. Two $Wx$ photonic crystal waveguides are fabricated in series. The width $D$ of the waveguide is adjusted to yield maximum transmission at the output of the coupler. Note that $D$ is assumed to be larger than the width of the input waveguide $w$ in order to provide rigid mechanical support and good clamping properties.

To overcome this difficulty we design supporting structures that provide a precise definition of the geometry of the final waveguide. Previously we have employed 1x1 multi-mode interference (MMI) couplers for this purpose. Even though MMI couplers can have low transmission loss, the achievable transmission is not sufficient to maintain the high optical Q of a ring or race-track resonator. We therefore employ a new design based on photonic crystal defect waveguides [14,15]. The principle is shown schematically in Fig.1.

*2.2 Photonic crystal coupler design*

We employ two-dimensional photonic crystals in a hexagonal lattice arrangement. The lattice constant $a$ of the photonic crystal and the hole diameter $h$ are adjusted to yield a wide bandgap around the operating wavelength of 1550nm in the telecoms C-band window. A defect waveguide is realized in the photonic crystal lattice by removing a number of rows along the direction of propagation. The resulting defect state in the bandgap of the bulk photonic crystal allows optical modes with wavelengths in the range of the defect state to propagate through the crystal lattice.

Traditionally the width $w$ of the incoming waveguide is chosen such that it equals the width of the defect waveguide, i.e. $w = \sqrt{3}a$, where $a$ is the lattice constant of the photonic crystal. This configuration is not suitable for our purposes because it leads to improper mechanical clamping at the edges of the removed holes. We therefore optimize the transmission of the coupler by requiring the width of the defect waveguide $D$ to be larger than $w+h$, where $h$ is the diameter of the air holes.

*2.2 Optimized parameters for high transmission through twin PhC couplers*

In order to optimize the design layout described above we employ finite-difference time-domain (FDTD) simulations using the commercial FDTD package CrystalWave. We assume a silicon-on-insulator (SOI) structure with a 3μm thick buried oxide layer and an 110nm thick top silicon layer. In order to achieve single mode propagation in the supporting waveguides the waveguide width $w$ is kept at 500nm. Throughout the paper we assume a refractive index of 3.477 for Silicon and 1.46 for the oxide layer. The photonic crystal region is assumed to be free-standing and is separated from the substrate by a 350nm wide air-gap.

We first optimize the parameters of the photonic crystal, i.e. the lattice constant $a$ and the hole diameter $h$, for high transmission in a standard W1 waveguide. Here the notation W1 denotes that one row of photonic crystal has been removed from the two-dimensional bulk to generate a defect waveguide [16]. For the 110nm thick silicon membrane optimal transmission is found for a lattice constant of 550nm and air hole diameter of 300nm.

The double-coupler structure is then improved for mechanical support by varying the PhC waveguide width *D*. We simulate the transmission through two PhC couplers joined by a mechanical beam for waveguide widths from 700nm to 1350nm. Each coupler is 17 lattice periods wide. The couplers are terminated at the center of a hole on both the input and output facets. Simulation results are shown in Fig.2a).

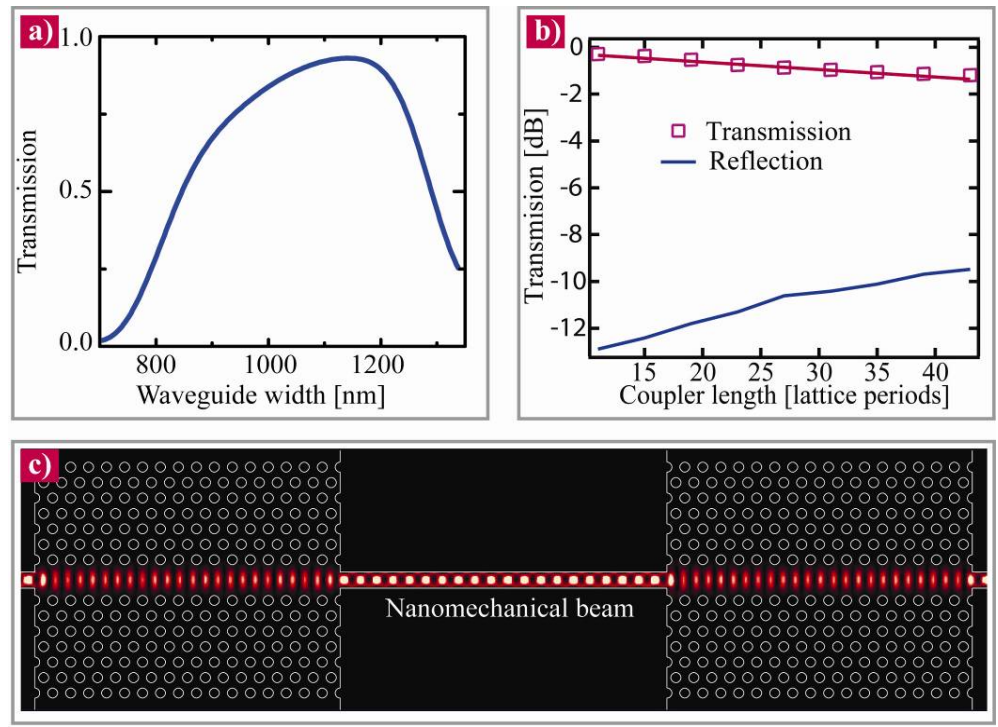

Figure 2: a) The transmission of the photonic crystal coupler in dependence of coupler width. Optimal transmission is found for a coupler width of 1150nm. The resulting transmission loss through a pair of couplers is less than -0.25dB. b) The dependence of the transmission on the width of the PhC couplers. Only a small decrease in transmission is observed over the investigated coupler lengths. c) The optical mode profile in the optimized coupler structure under illumination with CW light of wavelength 1550nm.

It is apparent, that the transmission depends strongly on the width of the PhC waveguide. When the width is properly chosen we find optimal transmission of 95%, corresponding to a loss of -0.23dB. This optimal width is found at 1150nm. When the width is either increased or decreased, the transmission drops sharply. However, we find that within tolerance band of 1150±50nm the transmission loss is less than 0.5dB and thus allows for potentially high optical quality factors. When the length of the couplers is varied we observe only small changes in the overall transmission. This is shown in Fig.2b), where we consider PhC couplers with lengths from 13 to 43 lattice periods. The width is kept constant at 1150nm. The transmission loss increases by -0.025dB with each additional lattice period. This is accompanied by an increase in reflection from the PhC Coupler. In this optimized structure the optical mode expands within the photonic crystal and is subsequently refocused on leaving the coupler as shown in Fig.2b). Minimal distortion of the optical mode is found for the optimal parameters given above, thus leading to low reflection from the PhC-beam interface.

*2.3 Investigation of the optical quality factor*

The optimized PhC coupler design is subsequently embedded in an optical race-track resonator. We simulate the entire design with the FDTD tool to obtain the optical quality factor of the resonator. We assume a resonator radius of 7.5μm and waveguide width of 500nm. The simulated radius of the race-track resonator is smaller than the fabricated device in the following section for computational efficiency. However, for a radius of 7.5μm the bend loss is negligible, therefore the quality factors of the resonator are expected to be similar. Reducing the radius will mostly increase the free spectral range of the resonator. As a control parameter we calculate the group index, which should be consistent with the experimentally determined value.

The thickness of the silicon layer is set to 110nm. The photonic crystal design parameters given above are employed for the coupling region. We assume that the photonic crystal has been released and is thus free-standing, separated by a gap of 350nm from the underlying substrate. The length of the nano-mechanical beam is 10μm. In order to obtain good extinction ratio in the transmission profile of the system, the race-track resonator is near-critical coupled by setting the distance between input waveguide and resonator to 200nm. The simulation is run for 50ps in order to resolve the frequency spectrum with good accuracy. Optical fields are recorded inside the race-track resonator and also on the far end of the input waveguide in order to allow for the extraction of the transmission data and the quality factor of the resonator. Results from the simulation are presented in Fig.3.

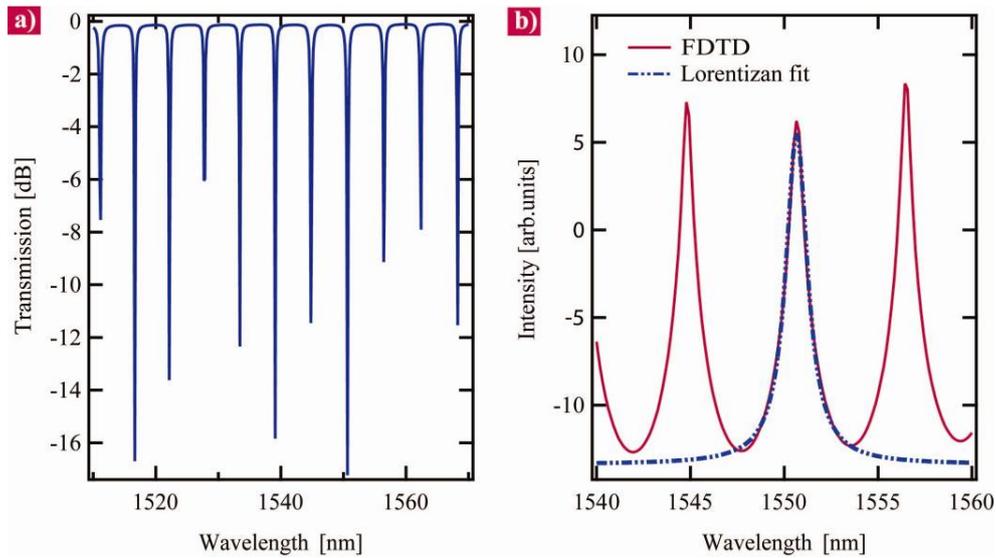

Fig.3 a) The transmission profile of a race-track resonator with a radius of 7.5μm. The coupling gap between the resonator and the input waveguide is 200nm. The free spectral range of the resonator is 5.6nm. b) The spectrum of the resonator inside the ring. The resonance at 1550nm is fitted with a Lorentzian and has a quality factor of ~3300.

In Fig.3a) we show the measured transmission profile of the resonator. We obtain an extinction ratio of roughly 17dB. This implies that most of the power coming from the input waveguide couples into the resonator, since bend losses are negligible in our case. We find a free spectral range (FSR) of 5.6nm, corresponding to a group index of the waveguide of 3.82. This value is consistent with control values obtained using finite-element simulations.

In order to measure the quality factor of the resonator we measure the spectrum of the intensity circulating inside the race-track. This is shown in Fig.3b). Fitting the resonant peaks in the spectrum with a Lorentzian function allow us to extract the optical quality factor. The depicted resonance at 1550nm shows a linewidth of 0.5nm and thus an optical quality factor of ~3300. Higher quality factors are found further away from the 1550nm resonance. In the wavelength range between 1500nm and 1600nm we find a peak quality factor of ~6000.

The results indicate that the transmission through the photonic crystal slab is indeed high. Therefore we can achieve good optical enhancement of the circulating power inside the race-track resonator which in turn leads to enhanced optical gradient forces.

## 3. Experiment

To verify the numerical predictions we fabricate nano-mechanical resonant devices inside a photonic circuit. We employ SOI wafers from Soitec, with an initial silicon thickness of 220nm. The top silicon layer is further thinned down to 110nm. Photonic circuitry is realized within the silicon layer by ebeam lithography and subsequent dry chemical etching. An optical micrograph of the fabricated circuit is shown in Fig.4a, showing the input waveguide and the race-track resonator. Light is coupled into the circuit using focusing grating couplers [17]. The grating couplers have a loss of roughly -10dB each, which could be reduced through optimized design. Nanowire waveguides of 500nm width are used to route light through the circuit. The chosen waveguide dimensions guarantee that only the fundamental mode in TE polarization is maintained in the nanowires. The waveguides typically have propagation loss of ~ 3dB/cm.

The race-track has a radius of 20μm to reduce bend losses to a minimum. The optical resonator is separated from the input waveguide by a gap of 200n, close to the critical coupling condition. For the photonic crystal couplers we use a lattice period of 550nm and a hole diameter of 300nm, following the results from the FDTD simulation. Each coupler is 17 lattice periods long and the PhC mirrors on each side of the defect waveguide are 11 rows wide. They are separated by a distance of 10μm to yield the nano-mechanical beam of desired length after release.

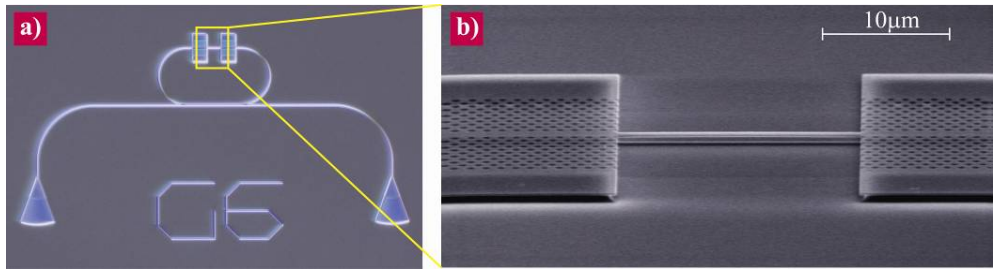

Fig.4 a) An optical micrograph of the photonic circuit used to measure both the optical and the mechanical quality factors. The image shows the circuit prior to the wet-chemical release step. b) An SEM image of the released part of the photonic circuit. A nano-mechanical beam resonator is realized by removing the underlying silicon dioxide layer.

Both the photonic crystal couplers and the beam between them are released from the substrate using wet-chemical etching with buffered oxide etchant. The etching release time is controlled such that a gap of 350nm is realized under the free-standing waveguide. An SEM image of the fabricated nano-mechanical device is shown in Fig.4b). The PhC couplers are supported on the sides where no air holes have been defined. The PhC itself is free-standing as well as the resonating beam. In this configuration sufficient optical waveguide gradient force can be generated to actuate and drive the nano-mechanical beam [18].

*3.1 Optical characterization of the photonic circuit*

To determine the optical quality factor of the race-track resonator we measure the transmission spectrum using a tunable diode laser and a photodetector. The transmission is recorded over the bandwidth of the input and output grating couplers, which provides a total observable bandwidth of roughly 40nm. The results are shown in Fig.5a).

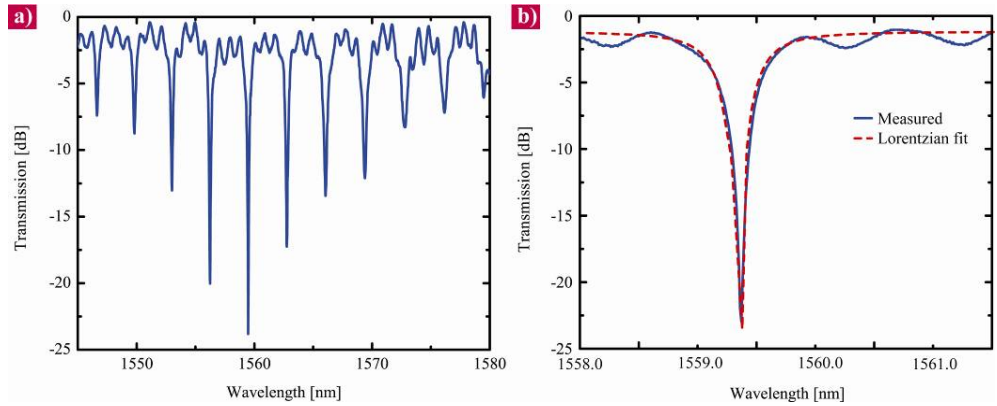

Fig.5 a) The normalized transmission spectrum of the race-track resonator. Resonances are observed over the coupling bandwidth of the input grating coupler with a FSR of 3.26nm. b) A close-up of the resonance with best extinction ratio. Fitting the dip with a Lorentzian allows us to extract the optical quality factor which is 4200 in this case.

Several resonances peaks are observed with a FSR of 3.26nm. The resulting group index is 3.87, in close agreement with the numerical prediction. Around 1560nm we find the optimal extinction ratio of almost 25dB. This illustrates that the resonator is almost critical coupled. In Fig.4b), a close-up of the resonance at 1559.4nm is shown, with a full-width half-minimum (FWHM) of 0.42nm and thus a quality factor of 4200. The Lorentzian fit is shown in the dashed line. The good quality factor confirms the high transmission of the PhC couplers.

*3.2 Nanomechanical characterization*

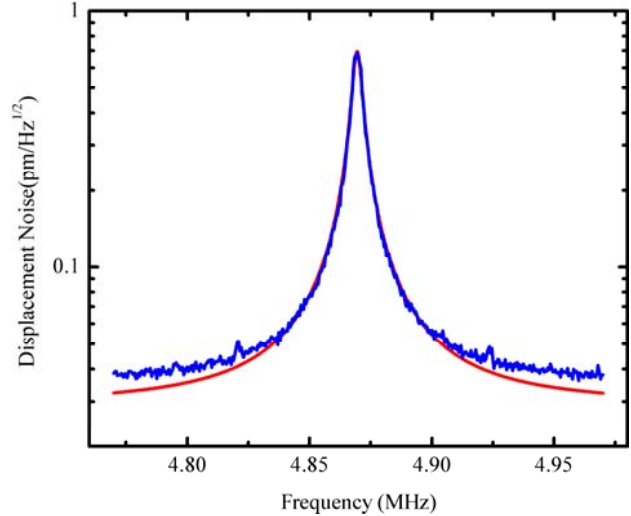

Fig.6 The mechanical response of the nano-resonator. The mechanical properties of the beam resonator are investigated by measuring the thermo-mechanical noise under vacuum. Shown are the measured signal (blue) and the fitted Lorentzian amplitude (dashed red line). The measured quality factor at a central frequency of 4.866 MHz is 1200.

Since the motion of the nanomechanical beam resonator modulates the effective refractive index of the waveguide and thus changes the resonance frequency of the ring resonator, the beam and the ring resonator are coupled dispersively. By applying a probing light with its frequency slightly detuned from the optical resonance frequency, the motion of the beam is read-out as amplitude modulation in the transmitted signal of the probe light. In Fig.6., we show the spectral power density (PSD) of the transmitted optical signal, measured at room temperature and vacuum pressure of 1 mTorr. The thermomechanical vibration of the beam is observed with a resonance peak at 4.88 MHz, which corresponds to the fundamental out-of plane motion of the beam. Fitting the resonance yields a mechanical quality factor of ~1200. This confirms the good mechanical clamping that is provided by the PhC couplers. The displacement detection resolution is 40 $fm/Hz^{1/2}$ which can be further improved by increasing the optical finesse of the ring resonator.

## 4. Conclusion

We have demonstrated both mechanical and optical enhancement of an optical-force generated signal. A high mechanical quality factor of 1200 and good optical Q-factors of ~4200 are obtained in a carefully designed photonic-crystal coupler supported race-track resonator. The PhC couplers provide good mechanical support without disturbing the propagating optical mode. In particular, this design allows for precise control of the length of the nano-mechanical resonator. In configurations without supporting couplers the beam length is determined by the undercut during the wet etching process. This procedure is susceptible to the experimental conditions and thus good repeatability is difficult to achieve. Our results show that both mechanical and optical signal enhancement are feasible in suitable cavity designs. We anticipate that the scheme proposed here will find applications in cavity optomechanical cooling of nano-mechanical devices.

**Acknowledgements**

We acknowledge support from DARPA/MTO. W.H.P. Pernice would like to thank the Alexander-von-Humboldt foundation for providing a postdoctoral fellowship. H.X. Tang acknowledges a Career award from National Science Foundation.